\documentclass[aps,prl,twocolumn,superscriptaddress]{revtex4}
\usepackage{graphicx}

\begin{document}

\title{Coherent dynamics of a flux qubit coupled to a harmonic oscillator}

\author{I. Chiorescu}\altaffiliation[Present address: ]{Department of Physics
and Astronomy, Michigan State University, East Lansing, Michigan 48824, USA} \affiliation{Quantum
Transport group, Kavli Institute of Nanoscience, Delft University of Technology, 2628 CJ, Delft,
Netherlands}
\author{P. Bertet}
\affiliation{Quantum Transport group, Kavli Institute of Nanoscience, Delft University of
Technology, 2628 CJ, Delft, Netherlands}
\author{K. Semba}
\affiliation{Quantum Transport group, Kavli Institute of Nanoscience, Delft University of
Technology, 2628 CJ, Delft, Netherlands} \affiliation{NTT Basic Research Laboratories, 3-1 Morinosato-Wakamiya,
Atsugi, 243-0198, Japan}
\author{Y. Nakamura} \affiliation{Quantum Transport group, Kavli Institute of Nanoscience, Delft University of
Technology, 2628 CJ, Delft, Netherlands} \affiliation{NEC Fundamental Research Laboratories, 34
Miyukigaoka, Tsukuba, Ibaraki 305-8501, Japan}
\author{C.J.P.M. Harmans}
\affiliation{Quantum Transport group, Kavli Institute of Nanoscience, Delft University of
Technology, 2628 CJ, Delft, Netherlands}
\author{J.E. Mooij}
\affiliation{Quantum Transport group, Kavli Institute of Nanoscience, Delft University of
Technology, 2628 CJ, Delft, Netherlands}

\date{Nature. Received 25 May; accepted 5 July 2004. doi:10.1038/nature02831 }

\maketitle

In the emerging field of quantum computation$^1$ and quantum information, superconducting devices
are promising candidates for the implementation of solid-state quantum bits or qubits.
Single-qubit operations$^{2-6}$, direct coupling between two qubits$^{7-10}$, and the realization of a
quantum gate$^{11}$ have been reported. However, complex manipulation of entangled states $-$ such as
the coupling of a two-level system to a quantum harmonic oscillator, as demonstrated in ion/atom-trap
experiments$^{12,13}$ or cavity quantum electrodynamics$^{14}$ $-$ has yet to be achieved for superconducting
devices. Here we demonstrate entanglement between a superconducting flux qubit (a two-level
system) and a superconducting quantum interference device (SQUID). The latter provides the measurement
system for detecting the quantum states; it is also an effective inductance that, in parallel with
an external shunt capacitance, acts as a harmonic oscillator. We achieve generation and control of the
entangled state by performing microwave spectroscopy and detecting the resultant Rabi oscillations
of the coupled system.

The device was realized by electron-beam lithography and metal evaporation. The qubit-SQUID
geometry is shown in Fig.~1a: a large loop interrupted by two Josephson junctions (the SQUID) is
merged with the smaller loop on the right-hand side comprising three in-line Josephson junctions
(the flux qubit)$^{15}$. By applying a perpendicular external magnetic field, the qubit is biased
around $\Phi_0/2$, where $\Phi_0=h/2e$ is the flux quantum. Previous spectroscopy$^{16}$ and
coherent time-domain experiments$^6$ have shown that the flux qubit is a controllable two-level
system with `spin-up/spin-down' states corresponding to persistent currents flowing in
`clockwise/anticlockwise' directions and coupled by tunneling. Here we show that a stronger
qubit$-$SQUID coupling allows us to investigate the coupled dynamics of a `qubit$-$harmonic
oscillator' system.

\begin{figure}
\begin{center}\includegraphics{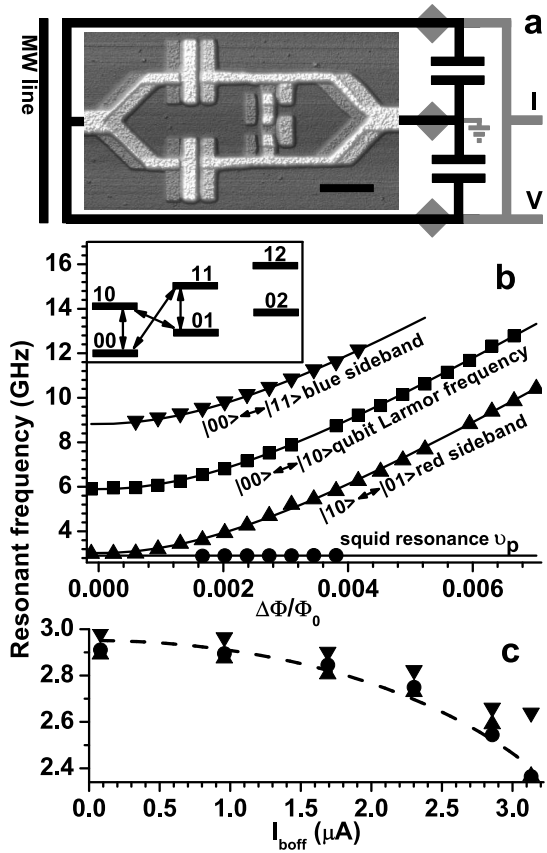}
\end{center}
\caption{Qubit$-$SQUID device and spectroscopy \textbf{a}, Atomic force micrograph
of the SQUID (large loop) merged with the flux qubit (the smallest loop closed by
three junctions); the qubit to SQUID area ratio is 0.37. Scale bar, $1 \mu m$. The SQUID (qubit) junctions have a
critical current of 4.2 (0.45)~$\mu$A. The device is made of aluminium by two symmetrically angled
evaporations with an oxidation step in between. The surrounding circuit shows aluminium shunt
capacitors and lines (in black) and gold quasiparticle traps$^3$ and resistive leads (in grey).
The microwave field is provided by the shortcut of a coplanar waveguide (MW line) and
couples inductively to the qubit. The current line ($I$) delivers the readout pulses, and the
switching event is detected on the voltage line ($V$). \textbf{b}, Resonant frequencies indicated by
peaks in the SQUID switching probability when a long microwave pulse excites the system before the readout
pulse. Data are represented as a function of the external flux
through the qubit area away from the qubit symmetry point. Inset, energy levels of the
qubit$-$oscillator system for some given bias point. The blue and red sidebands are shown by down-
and up-triangles, respectively; continuous lines are obtained by adding $2.96$~GHz and $-2.90$~GHz,
respectively, to the central continuous line (numerical fit). These values are close to the oscillator resonance $\nu_p$
at 2.91 GHz (solid circles) and we attribute the small differences to the slight dependence of $\nu_p$
on qubit state. \textbf{c}, The plasma resonance (circles) and the distances between the qubit peak
(here $F_L=6.4$~GHz) and the red/blue (up/down triangles) sidebands as a function of an offset
current $I_{boff}$ through the SQUID. The data are close to each other and agree well with the
theoretical prediction for $\nu_p$ versus offset current (dashed line).}
\end{figure}

\begin{figure}
\begin{center}\includegraphics{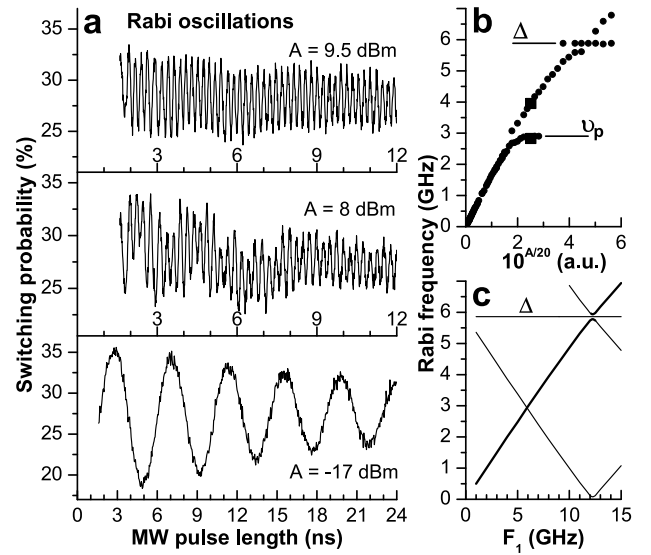}
\end{center}
\caption{Rabi oscillations at the qubit symmetry point $\Delta=5.9$~GHz. \textbf{a}, Switching
probability as a function of the microwave pulse length for three microwave nominal powers; decay
times are of the order of 25 ns. For $A=8$~dBm, bi-modal beatings are visible (the corresponding
frequencies are shown by the filled squares in \textbf{b}). \textbf{b}, Rabi frequency, obtained
by fast Fourier transformation of the corresponding oscillations, versus microwave amplitude.
In the weak driving regime, the linear dependence is in agreement with estimations based on sample
design. A first splitting appears when the Rabi frequency is $\sim\!\!\nu_p$. In the strong driving regime,
the power independent Larmor precession at frequency $\Delta$ gives rise to a second splitting. \textbf{c}, This
last aspect is obtained in numerical simulations where the microwave driving is represented by a term
$(1/2)hF_1\cos(\Delta t)$ and a small deviation from the symmetry point (100 MHz) is introduced in
the strong driving regime (the thick line indicates the main Fourier peaks).
Radiative shifts$^{20}$ at high microwave power could account for such a shift
in the experiment. }
\end{figure}

\begin{figure}
\begin{center}\includegraphics{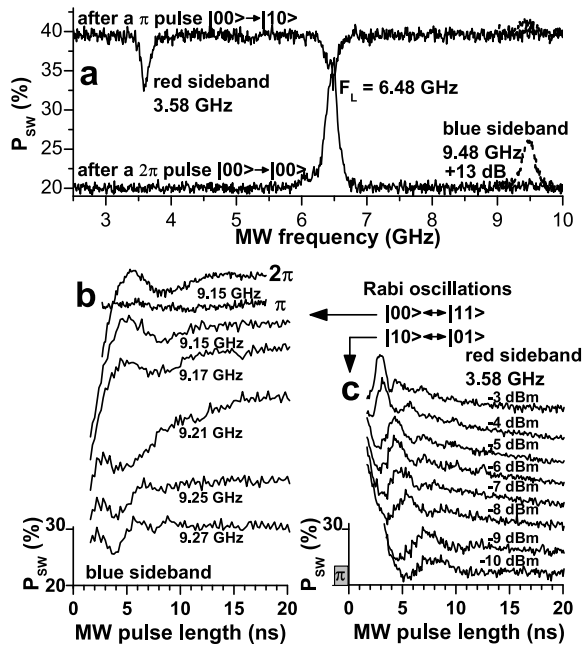}
\end{center}
\caption{Generation and control of entangled states. \textbf{a}, Spectroscopic characterization of
the energy levels (see Fig.~1b inset) after a $\pi$ (upper scan) and a $2\pi$ (lower scan) Rabi
pulse on the qubit transition. In the upper scan, the system is first excited to $|10\rangle$
from which it decays towards the $|01\rangle$ excited state (red sideband at 3.58~GHz) or towards the
$|00\rangle$ ground state ($F_L=6.48$~GHz). In the lower scan, the system is rotated back to the
initial state $|00\rangle$ wherefrom it is excited into the $|10\rangle$ or $|11\rangle$ states (see, in
dashed, the blue sideband peak at 9.48~GHz for 13~dB more power). \textbf{b}, Coupled Rabi
oscillations: the blue sideband is excited and the switching probability is recorded as a function
of the pulse length for different microwave powers (plots are shifted vertically for clarity). For
large microwave powers, the resonance peak of the blue sideband is shifted to 9.15~GHz. When
detuning the microwave excitation away from resonance, the Rabi oscillations become faster (bottom
four curves). These oscillations are suppressed by preparing the system in the $|10\rangle$ state
with a $\pi$ pulse and revived after a $2\pi$ pulse (top two curves in Fig. 3b) \textbf{c},
Coupled Rabi oscillations: after a $\pi$ pulse on the qubit resonance ($|00\rangle\rightarrow|10\rangle$)
we excite the red sideband at 3.58~GHz. The switching probability shows coherent oscillations
between the states $|10\rangle$ and $|01\rangle$, at various microwave powers (the curves are shifted
vertically for clarity). The decay time of the coherent oscillations in \textbf{a}, \textbf{b}
is $\sim\!\!3$~ns. }
\end{figure}

\begin{figure}
\begin{center}\includegraphics{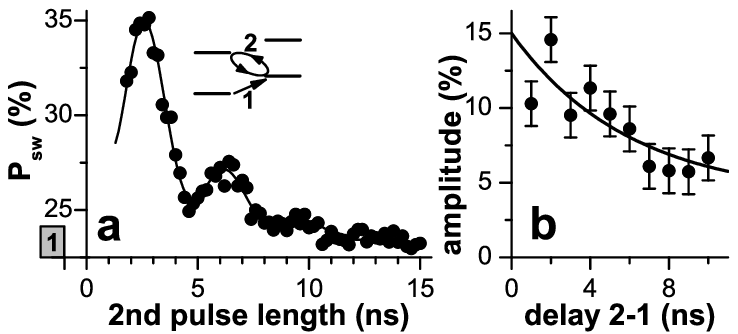}
\end{center}
\caption{Oscillator relaxation time. \textbf{a}, Rabi oscillations between the $|01\rangle$ and
$|10\rangle$ states (during pulse 2 in the inset) obtained after applying a first pulse (1) in
resonance with the oscillator transition. Here, the interval between the two pulses is 1 ns.
The continuous line represents a fit using an exponentially decaying sinusoidal oscillation
plus an exponential decay of the background (due to the relaxation into the ground state).
The oscillation's decay time is $\tau_{coh}=2.9$~ns, whereas the background decay time is $\sim4$~ns.
\textbf{b}, The amplitude of Rabi oscillations as a function of the interval
between the two pulses (the vertical bars represent standard error bars estimated from the fitting procedure,
see \textbf{a}). Owing to the oscillator relaxation, the amplitude decays in $\tau_{rel}\approx6$~ns
(the continuous line represents an exponential fit).}
\end{figure}

The qubit Hamiltonian is defined by the charging and Josephson energy of the qubit outer junctions
($E_C=e^2/2C$ and $E_J=hI_C/4e$ where $C$ and $I_C$ are their capacitance and critical
current)$^{16}$. In a two-level truncation, the Hamiltonian becomes
$H_q/h=-\epsilon\sigma_z/2-\Delta\sigma_x/2$ where $\sigma_{z,x}$ are the Pauli matrices in the
spin-up/spin-down basis, $\Delta$ is the tunnel splitting and $\epsilon\cong
I_p\Phi_0(\gamma_q-\pi)/h\pi$ ($I_p$ is the qubit maximum persistent current and $\gamma_q$ is the
superconductor phase across the three junctions). The resulting energy level spacing represents
the qubit Larmor frequency $F_L=\sqrt{\Delta^2+\epsilon^2}$. The SQUID dynamics is characterized
by the Josephson inductance of the junctions $L_J\approx80$~pH, shunt capacitance $C_{sh}\approx12$~pF
(see Fig.~1a) and self-inductances $L_{sl}\approx170$~pH of the SQUID and shunt-lines. In our
experiments, the SQUID circuit behaves like a harmonic oscillator described by
$H_{sq}=h\nu_p(a^\dag a+1/2)$, where $2\pi\nu_p=1/\sqrt{(L_J+L_{sl})C_{sh}}$ is called the plasma
frequency and $a$ ($a^\dag$) is the plasmon annihilation (creation) operator. Henceforth
$|\beta n\rangle$ represents the state with the qubit in the ground($\beta=0$) or excited
($\beta=1$) level, and the oscillator on the $n^{th}$ level
($n=0,1,2,\ldots$). The corresponding level diagram is sketched in Fig.~1b (inset). The coupling
between the qubit and the oscillator originates from the current distribution in the shared
branches (Fig.~1a) and gives rise to an interaction Hamiltonian
$H_{q-sq}=\lambda\sigma_z(a+a^\dag)$ with $\lambda\approx0.2$~GHz in our device$^{17}$
(the estimated qubit-SQUID coupling is $M\approx20$~pH).

Measurements are performed at T=25~mK using low-noise circuitry to minimize decoherence,
relaxation and thermal activation. The system is first initialized by allowing it to relax to the
$|00\rangle$ ground state. With successive resonant microwave pulses we achieve controlled
superposition of various $|\beta n\rangle$ states, as shown below. The readout$^6$ is performed by
applying a short current pulse $I_b$ ($\sim\!\!30$~ns) and by monitoring whether the SQUID
switches to the finite-voltage state. After averaging ty--pically 10000 readouts, we obtain the
probability $P_{sw}$($I_b$) which for properly-chosen parameters is proportional to the excited
state occupancy. In the following, we first show the spectroscopy of the coupled
qubit$-$oscillator system and Rabi oscillations of the qubit. Next we demonstrate coherent
dynamics of the coupled system.

We performed spectroscopy of the coupled qubit$-$oscillator system by applying a long (300~ns)
microwave pulse with various frequencies and measuring the SQUID switching probability. Peaks and
dips are observed and their resonant frequencies as a function of $\Delta\Phi=\Phi_{ext}-\Phi_0/2$
are given in Fig.~1b. We obtain one manifold of three resonances spaced by $\sim\!\!2.91$~GHz.
This frequency coincides with the designed oscillator eigenfrequency $\nu_p$. In addition, we
observe a spectroscopic peak or dip that depends only weakly on the magnetic field (circles in
Fig. 1b). For lower microwave power, only the qubit band (squares) remains visible. A numerical
fit (continuous line) of this band leads to $E_J=225$~GHz, $E_C=7.3$~GHz, and ratio of area of qubit junctions
$\alpha=0.76$ ($\Delta=5.9$~GHz, $I_p=275$~nA). The appearance of the manifold instead of a single
resonance is due to the qubit coupling with the oscillator mode $\nu_p$(ref. $18$). Similarly to atomic
physics, we call the $|00\rangle$ to $|11\rangle$ ($|01\rangle$ to $|10\rangle$) transitions the blue
(red) sidebands (see the ladder energy diagram of the $|\beta n\rangle$ states in Fig.~1b inset).
We note that near the qubit symmetry point, the closeness of the oscillator resonance and the red
sideband, visible owing to a small thermal occupation of the $|01\rangle$ state, is purely accidental.
To verify that the oscillator involved is indeed the SQUID plasma mode, we repeated the above
measurements in the presence of an offset bias current $I_{boff}$ which decreases the plasma frequency
following$^{19}$ $\nu_p(1-(I_{boff}/I_c)^2)^{1/4}$, where $I_c$ is the SQUID critical current (4.2
A). The data in Fig.~1c show the distance between the qubit peak for $F_L=6.4$~GHz and the
blue/red sidebands (down/up triangles) that decreases together with the oscillator resonance
(circles).

To realize quantum operations on the qubit only, we apply a resonant microwave pulse with
frequency $F_{mw} = F_L$. The operation is performed at the qubit symmetry point $\gamma_q=\pi$
where $F_L=\Delta$. In Fig. 2a, the SQUID switching probability is plotted against the microwave
pulse length for three microwave power levels. The observed Rabi oscillations decay within
$\sim\!\!30$~ns. Remarkably, we can reach Rabi frequencies comparable to the Larmor frequency (up
to 6.6~GHz). Using Fourier transformation, we extract the Rabi frequency as a function of the
microwave amplitude (Fig.~2b). In the weak driving regime, the Rabi frequency increases linearly
with the microwave power, as expected$^6$. Near the oscillator resonance $\nu_p$, we see two
frequencies in the spectrum, a behaviour which is probably caused by the qubit$-$oscillator
coupling. At even higher microwave powers, the spectrum exhibits again a second frequency
component at $\Delta$. A qualitatively similar behaviour is also obtained in numerical simulations
(see Fig.~2c) when we consider the qubit driven by an additional term $(1/2)hF_1\cos(\Delta t)$ in
$H_q$ ($F_1$ and $\Delta$ are the microwave amplitude and frequency, respectively).

We now turn to the conditional dynamics resulting from the qubit$-$oscillator coupling. We first
determine the blue and red sideband resonant frequencies by spectroscopic means using a two-pulse
sequence (Fig.~3a). The qubit is prepared in the excited state by a $\pi$ pulse at the Larmor
frequency. A second pulse (18~ns) of variable frequency induces resonant qubit de-excitation (dips in
Fig.~3a top trace) marking the red sideband and the Larmor frequency. Similarly, after a $2\pi$ pulse
which places the qubit in its ground state, we search for resonant excitations (peaks in Fig.~3a
bottom trace) that mark the Larmor frequency and the blue sideband. No resonance is seen on the
red sideband, showing that the oscillator is in its ground state with a large probability. Note
that in order to excite the blue sideband, we have to increase the microwave power by at least
$10$~dB, probably due to less effective microwave transmission in the $8-9$~GHz range (note also
the absence of spectroscopy peaks in this frequency range in Fig.~1b). At high microwave powers,
we observe radiative shifts$^{20}$ of the resonances. We now exploit these resonances to study the
dynamics of the coupled system by applying pulses of varying length. In Fig.~3b, Rabi oscillations
are shown for the $|00\rangle$ to $|11\rangle$ transition. When the microwave frequency is detuned
from resonance, the Rabi oscillations are accelerated (bottom four curves, to be compared with the
fifth curve). After a $\pi$ pulse which prepares the system in the $|10\rangle$ state, these
oscillations are suppressed (second curve in Fig. 3b). After a $2\pi$ pulse they are revived
(first curve in Fig.~3b). In the case of Fig.~3c, the qubit is first excited onto the $|10\rangle$
state by a $\pi$ pulse and a second pulse in resonance with the red sideband transition drives the
system between the $|10\rangle$ and $|01\rangle$ states. The Rabi frequency depends linearly on the
microwave amplitude, with a smaller slope compared to the bare qubit driving. During the time
evolution of the coupled Rabi oscillations shown in Figs.~3b and 3c, the qubit and the oscillator
experience a time-dependent entanglement, although the present data do not permit us to quantify
it to a sufficient degree of confidence.

The sideband Rabi oscillations of Fig.~3 show a short coherence time ($\sim\!\!3$~ns) which we
attribute mostly to the oscillator relaxation. To determine its relaxation time, we performed the
following experiment. First, we excite the oscillator with a resonant low power microwave pulse.
After a variable delay $\Delta t$, during which the oscillator relaxes towards $n=0$, we start
recording Rabi oscillations on the red sideband transition (see Fig.~4a for $\Delta t=1$~ns). The
decay of the oscillation amplitude as a function of $\Delta t$ corresponds to an oscillator
relaxation time of $\sim\!\!6$~ns (Fig.~4b), consistent with a quality factor of $100-150$
estimated from the width of the $\nu_p$ resonance. The exponential fit (continuous line in
Fig.~4b) shows an offset of $\sim\!\!4\%$ due to thermal effects. To estimate the higher bound of the sample
temperature, we consider that the visibility of the oscillations presented here
(Figs.~2-4) is set by the detection efficiency and not by the state preparation. When related to
the maximum signal of the qubit Rabi oscillations of $\sim\!\!40\%$, the $4\%$-offset corresponds
to $\sim\!\!10\%$ thermal occupation of oscillator excited states (an effective temperature of
$\sim\!\!60$~mK). Consistently, we also observe low-amplitude red sideband oscillations without
preliminary microwave excitation of the oscillator.

We have demonstrated coherent dynamics of a coupled superconducting two-level plus
harmonic oscillator system, implying that the two subsystems are entangled. Increasing the
coupling strength and the oscillator relaxation time should allow us to quantify the entanglement,
as well as to study non-classical states of the oscillator. Our results provide strong
indications that solid-state quantum devices could in future be used as elements for the
manipulation of quantum information.

\begin{acknowledgments}
We thank A. Blais, G. Burkard, D. DiVincenzo, G. Falci, M. Grifoni, S. Lloyd, S. Miyashita, T. Orlando, R. N.
Schouten, L. Vandersyepen, F. K. Wilhelm for discussions. This work was supported by the
Dutch Foundation for Fundamental Research on Matter (FOM), the E.U. Marie Curie and SQUBIT grants,
and the U.S. Army Research Office. The authors declare that they have no competing financial
interests. Correspondence and requests for materials should be addressed to I.C. (e-mail:
chiorescu@pa.msu.edu) and J.E.M. (email: mooij@qt.tn.tudelft.nl).
\end{acknowledgments}


\begin{thebibliography}{99}

\bibitem{ref01} Nielsen, M. A., Chuang, I. L. \emph{Quantum Computation and Quantum Information}
(Cambridge Univ. Press, Cambridge, 2000).

\bibitem{ref02} Nakamura, Y. \emph{et al.} Coherent control of macroscopic quantum states in a single-Cooper-pair
box. \emph{Nature} \textbf{398}, 786-788 (1999).

\bibitem{ref03} Vion, D. \emph{et al.} Manipulating the quantum state of an electrical circuit. \emph{Science} \textbf{296}, 886 889
(2002).

\bibitem{ref04} Yu, Y., Han, S., Chu, X., Chu, S., Wang, Z. Coherent temporal oscillations of macroscopic
quantum states in a Josephson junction. \emph{Science} \textbf{29}6, 889-892 (2002).

\bibitem{ref05} Martinis, J. M., Nam, S., Aumentado, J., Urbina, C., Rabi oscillations in a large
Josephson-junction qubit. \emph{Phys. Rev. Lett.} \textbf{89}, 117901 (2002).

\bibitem{ref06} Chiorescu, I., Nakamura, Y., Harmans, C. J. P. M., Mooij, J. E. Coherent quantum dynamics of a
superconducting flux qubit. \emph{Science} \textbf{299}, 1869-1871 (2003).

\bibitem{ref07} Pashkin, Yu. A., \emph{et al.} Quantum oscillations in two coupled charge qubits. \emph{Nature} \textbf{421}, 823-826
(2003).

\bibitem{ref08} Berkley, A. J., \emph{et al.} Entangled macroscopic quantum states in two superconducting qubits.
\emph{Science} \textbf{300}, 1548-1550 (2003).

\bibitem{ref09} Majer, J. B., Paauw, F. G., ter Haar, A. C. J., Harmans, C. J. P. M., Mooij, J. E.
Spectroscopy on two coupled flux qubits. Preprint at $<$http://arxiv.org/abs/cond-mat/0308192$>$ (2003).

\bibitem{ref10} Izmalkov, A., \emph{et al.} Experimental evidence for entangled states formation in a system of two
coupled flux qubits. Preprint at $<$http://arxiv.org/abs/cond-mat/0312332$>$ (2003).

\bibitem{ref11} Yamamoto, T., Pashkin, Yu. A., Astafiev, O., Nakamura, Y., Tsai, J. S. Demonstration of
conditional gate operation using superconducting charge qubits. \emph{Nature} \textbf{425}, 941-944 (2003).

\bibitem{ref12} Leibfried, D., Blatt, R., Monroe, C., Wineland, D., Quantum dynamics of single trapped ions.
\emph{Rev. Mod. Phys.} \textbf{75}, 281-324 (2003).

\bibitem{ref13} Mandel, O., \emph{et al.} Controlled collisions for multi-particle entanglement of optically trapped
atoms. \emph{Nature} \textbf{425}, 937-940 (2003).

\bibitem{ref14} Raimond, J.M., Brune, M., Haroche, S. Manipulating quantum entanglement with atoms and photons
in a cavity. \emph{Rev. Mod. Phys.} \textbf{73}, 565-582 (2001).

\bibitem{ref15} Mooij, J. E., \emph{et al.} Josephson Persistent-Current Qubit. \emph{Science} \textbf{285}, 1036-1039 (1999).

\bibitem{ref16} van der Wal, C. H., \emph{et al.} Quantum superposition of macroscopic persistent-current states.
\emph{Science} \textbf{290}, 773-777 (2000).

\bibitem{ref17} Burkard, G., \emph{et al.} Asymmetry and decoherence in double-layer persistent-current qubit.
Preprint at $<$http://arxiv.org/abs/cond-mat/0405273$>$ (2004).

\bibitem{ref18} Goorden, M.C., Thorwart, M., Grifoni, M. Entanglement spectroscopy of a driven solid-state
qubit and its detector. Preprint at $<$http://arxiv.org/abs/cond-mat/0405220$>$ (2004).

\bibitem{ref19} Tinkham, M., \emph{Introduction to superconductivity} 2nd edn (McGraw-Hill, New York, 1996), pp. 207.

\bibitem{ref20} Cohen-Tannoudji, C., Dupont-Roc, J., Grynberg, G. \emph{Atom-photon interactions: basic processes
and applications} Ch. II E (John Wiley \& Sons, New York, 1992).

\end{thebibliography}
\end{document}